\documentstyle[aps,prb,floats]{revtex}

\begin{document}
\input epsf
\draft
\renewcommand{\floatpagefraction}{0.99}
\renewcommand{\topfraction}{0.99}
\twocolumn[\hsize\textwidth\columnwidth\hsize\csname@twocolumnfalse%
\endcsname

\title{  {\rm\small\hfill submitted to PRB, 10/28/2000}\\
Surface Core Level Shifts of Clean and Oxygen Covered Ru(0001)}

\author{S. Lizzit, A. Baraldi, A. Groso}
\address{$^1$Sincrotrone Trieste S.C.p.A., S.S. 14 Km 163.5, 34012 Trieste, Italy}

\author{K. Reuter, M.V. Ganduglia-Pirovano, C. Stampfl, M. Scheffler}
\address{$^2$Fritz-Haber-Institut der Max-Planck-Gesellschaft, Faradayweg
4-6, D-14195 Berlin-Dahlem, Germany}

\author{M. Stichler, C. Keller, W. Wurth, and D. Menzel}
\address{$^3$Physik-Department E20, Techn. Universit\"at M\"unchen,
D-85748 Garching, Germany }

\date{\today}
\maketitle

\begin{abstract}
We have performed high resolution XPS experiments of the Ru(0001) surface,
both clean and covered with well-defined amounts of oxygen up to 1 
ML coverage. For the clean surface we detected two distinct components in 
the Ru $3d_{5/2}$ core level spectra, for which a definite assignment was
made using the high resolution Angle-Scan Photoelectron Diffraction approach.
For the $p(2\times 2)$, $p(2\times 1)$, $(2\times 2)$-3O and $(1\times 1)$-O
oxygen structures we found Ru $3d_{5/2}$ core level peaks which are shifted
up to 1 eV to higher binding energies. Very good agreement with density
functional theory calculations of these Surface Core Level Shifts (SCLS) is
reported. The overriding parameter for the resulting Ru SCLSs turns out to
be the number of directly coordinated O atoms. Since the calculations permit
the separation of initial and final state effects, our results give
valuable information for the understanding of bonding and screening at the
surface, otherwise not accessible in the measurement of the core level
energies alone.
\end{abstract}

\hfill {\quad}

]

\section{Introduction} 

The interaction of oxygen with transition metal surfaces is of considerable
interest. Apart from its model character for adsorbate-substrate interactions,
it is important as the first step of oxidation of these surfaces, and because
of its involvement in catalytic reactions such as CO oxidation, used e.g.
for the decontamination of automobile exhaust gases. Therefore, significant
efforts have been made in the last decades to investigate this model process,
both from an experimental and theoretical point of view. Oxygen chemisorption
on transition metal surfaces is largely discussed in terms of strong covalent
bonding between the O $2p$ states and the metal valence $d$-band, accompanied
by an unspecified, but noticeable charge transfer from the substrate to the
electronegative adsorbate. However, it is not clear which part of the
total electron density could or should be assigned to which atom, so that a
clearcut distinction between charge transfer and polarization is not possible
\cite{bormet94,scheffler00}.

In this context, theoretical concepts have been developed that try to
partition a calculated total electron density into contributions from
individual atoms \cite{mulliken55,hoffmann88,bader90,becke90}. Yet, it
would also be useful to have an experimentally accesible quantity, which
gives information about the nature of the chemical bond or which would
even help to quantify the amount of charge transferred. As core levels
are relatively compact and are generally assumed not to take part in
the bonding itself, core level binding energies provide such a local
probe of the changes in the electrostatic potential of an atom in different
environments. At surfaces, the core level energies of the substrate
atoms are changed relative to the bulk, giving rise to the so-called
Surface Core Level Shifts (SCLS), which can be measured both
for clean and adsorbate covered surfaces by high resolution core
level photoemission spectroscopy \cite{martensson94,baraldi00a}. However,
total SCLSs comprise not only the so-called initial state
effects, which reflect the changes in the electronic
distribution at the unperturbed surface, i.e. before the excitation of
the core hole, but also the final state effects which are due to
the different screening capabilities of the already core-ionized system
at the surface and in the bulk \cite{spanjaard85}. 
Here, a complementary analysis by density functional theory (DFT) is
important, because the latter is able to subdivide the total SCLSs
into initial and final state contributions. 

Such an approach of coupling experiment and theory has already been
used to study the SCLSs of clean transition metals \cite{andersen94},
and recently also to analyse adsorbate induced SCLSs due to the
interaction of O with the Rh(111) surface for the $p(2 \times 2)$ and
$p(2 \times 1)$ ordered adlayer structures \cite{ganduglia00}.
The present investigation of the O interaction with the
Ru(0001) surface aims to compare the chemisorption behaviour of the
two surfaces. Further, on Ru(0001) four different ordered O adlayer
structures are formed, which span the coverage range from zero up to
one monolayer (ML) and are all extensively characterized by LEED
experiments \cite{lindroos89,pfnuer89,gsell98,stampfl96a} and DFT
calculations \cite{stampfl96b}. Hence, a much larger experimental
data base is available compared to the O/Rh(111) work, which allows 
to assess much better the agreement between measured and calculated SCLSs.
The four ordered oxygen overlayers, which we have prepared and studied
besides the clean surface, are the $p(2 \times 2)$ \cite{lindroos89},
the $p(2 \times 1)$ \cite{pfnuer89}, the $(2 \times 2)$-3O \cite{gsell98}
and the $(1 \times 1)$-O \cite{stampfl96a} structure. In all phases, the
O atoms sit in hcp hollow sites and the Ru atoms can have up to three
O neighbours as shown in Fig. 1.

\begin{figure}
       \epsfxsize=0.47\textwidth \centerline{\epsfbox{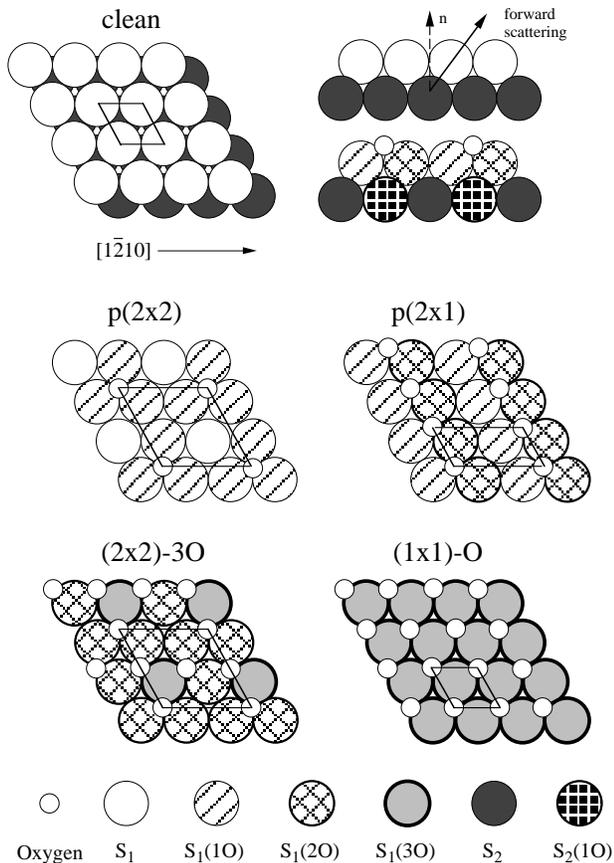}}
       \caption{Periodic oxygen adlayer structures on the Ru(0001) surface
        with increasing coverage. $S_1$, $S_1$(1O), $S_1$(2O)
        and $S_1$(3O) are first layer Ru atoms 
        bound to no, one, two, and three oxygen
        atoms, respectively. $S_2$ and $S_2$(1O) are second layer atoms 
        with no and one oxygen atom directly above on the surface, 
        respectively. The bulk $b$ includes all deeper layer Ru atoms.
        The top right panel shows sideviews of the clean
        Ru(0001) surface with an indication of the angle at which strong forward
        scattering is expected, and of the $p(2\times 1)$ structure.
        \label{fig:1}}
\end{figure}

As will be shown in section IV A, the Ru $3d_{5/2}$
core level spectra are composed of several peaks, which have to be
assigned to certain bonding situations of the corresponding Ru atoms.
From the aforementioned work on O/Rh(111), we expect the SCLSs
of the first layer atoms to depend primarily on the number of directly
coordinated O atoms. The nomenclature that we use to name each of
these atoms (and their corresponding SCLS) is derived from this fact 
and is described in Fig. 1.

If the number of nearest neighbour O atoms is indeed the ruling quantity for
the first layer peaks, the assignment of the O-induced components in the
spectra is straightforward, because each such peak should be present
in two of the considered phases. As shown
in Fig. 2, O-induced components at approximately equal positions
appear indeed each time at two coverages, so that recurrently working
down from the $(1 \times 1)$-O, the $S_1(\rm{3O})$, $S_1(\rm{2O})$,
and $S_1(\rm{1O})$ peaks can directly be assigned. Unfortunately, the
situation is not so simple for the $S_1$ and $S_2$ peaks, which are
both present in the spectrum of the clean surface and of the 
$p(2\times 2)$ phase. While the favorable comparison of experiment and
theory to be reported in the present work does also offer an assignment
for these peaks, it is still desirable to reach assignments on experimental
grounds only. In previous works, high resolution Photoelectron Diffraction
in the forward scattering regime had already been successfully utilized
to assign different components to first and second layer atoms
\cite{lizzit98,baraldi00}. In this work, a similar strategy will be pursued
for the clean Ru(0001) surface, in order to independently assign
the remaining $S_1$ and $S_2$ components. Once the measurement and the
assignment of the various SCLS components has been accomplished, they
can be compared with the theoretical results. As the latter allow to
separate the final state contribution from the total shift, we are then
in a position to discuss the connection of the initial state shift with
the nature of the chemical bond.

\section{Experimental}

The SCLS experiments were performed at the superESCA beamline of the ELETTRA
syncrotron facility in Trieste, Italy \cite{baraldi95}. The experimental
chamber is equipped with a new double pass electron energy analyser (which is 
composed of two hemispheres of 150 mm radius each) \cite{baraldi94} with a
96-channels detector \cite{gori99} (some earlier results were obtained with a
VSW spherical analyser), a VG manipulator (CTPO) with five degrees of freedom
and with heating and cooling capabilities (1500\ K and 120\ K, respectively),
a Leybold rear view LEED optics, and a channelplate doser for dosing high
amounts of oxygen. All data shown for the series of SCLSs as a function of
oxygen coverage were measured in one single run for maximum comparability, but
were in good agreement with a partial data set obtained earlier using a VSW
150 mm electron energy analyser with 16 channels parallel detection
\cite{stichler}. The photoelectron diffraction experiment was carried out in a
separate run, also using the VSW analyser.

The Ru(0001) crystal was cleaned by Ar$^{+}$ sputtering and repeated cycles
of oxygen treatments at temperatures ranging from 1000\ K to 1200\ K. Finally,
the sample was flashed to 1500\ K and cooled down to 300\ K in
$1\times10^{-7}$ mbar hydrogen pressure in order to remove any residual trace
of oxygen; to remove the hydrogen, the sample was briefly heated to 500\ K in
UHV before measurements. A very sharp $(1 \times 1)$ LEED pattern with low
background intensity was obtained and the XPS did not show any trace of
carbon, oxygen, or other contaminants.

The SCLS spectra, both in the measurement of the oxygen structures and in the
diffraction experiment, were acquired at a sample temperature lower than 130\
K and at a base pressure of $6 \times 10^{-11}$ mbar. Before doing the SCLS
measurements, the different oxygen structures were defined by observing the
intensity of the ($1\over2$,$1\over2$) spot in the LEED pattern induced by the
oxygen adsorption. The fully developed three structures up to 0.75 ML show
maxima in the intensity of the extra spots while dosing oxygen when the layer
corresponds to 0.25, 0.50 and 0.75 ML coverage. Since the LEED apparatus is
mounted in the experimental chamber we could in this way monitor the correct
dose of oxygen in order to obtain the desired structure. The coverage was also
checked by measuring the O$_{1s}$ intensity. Comparison of the LEED to the XPS
data shows that the O$_{1s}$ signal measured at 650 eV photon energy is not
much affected by diffraction effects; it therefore gives a good estimate of
the relative coverage.

The $p(2\times 2)$ structure was obtained by exposing the clean Ru(0001)
surface to 0.7 Langmuir (nominal) of oxygen at 373\ K, and subsequent brief
heating to 670\ K. The $p(2\times 1)$ structure was obtained by dosing onto
the $p(2\times 2)$ additional 3.5 Langmuir at 373\ K, followed again by brief
heating to 670\ K. As reported in the previous works, flashing at 670\ K after
the doses is needed to achieve perfect order of the superstructure. The
$(2\times 2)$-3O structure was obtained by dosing oxygen for 600 seconds with
the channel plate doser at a distance of ~10 mm from the sample, with a
pressure in the chamber of $1.5\times10^{-6}$ mbar at a sample temperature 
of 600\ K. The resulting O$_{1s}$ intensity corresponded to 0.85 ML. In order
to remove the excess oxygen the sample was briefly heated to 1060\ K; the
resulting coverage was 0.77 ML. The $(1\times 1)$-O structure was obtained by
dosing NO$_2$ 3 times, 800 seconds each, with the doser (pressure in the
chamber $5\times10^{-8}$ mbar), at a sample temperature of 600\ K. A very
sharp $(1 \times 1)$ LEED pattern resulted.

The high resolution Ru $3d_{5/2}$ SCLS spectra were recorded at a photon
beam incidence angle of $80^{\circ}$ from the surface normal; in the used
machine this leads to an electron emission angle of $40^{\circ}$.
Three different photon energies, 352, 370 and 400 eV were used in order to
change the weight of the core level components due to diffraction and inelastic
scattering effects. The $p(2\times 2)$ structure was measured only at 
352 eV. The analyser was operated at 5 eV pass energy with an entrance 
slit of 2 mm. The combined (photon plus electron) energy resolution is
estimated to have been better than 80 meV. For the photoelectron diffraction
measurements on the clean Ru(0001) surface, we used a photon energy of 500 eV,
which corresponds to a kinetic energy of the Ru $3d_{5/2}$ core level of 220
eV, high enough to have strong forward scattering effects. We performed an
azimuthal scan at $40^{\circ}$ emission angle with the photon beam now
parallel to the surface normal. Since at this high photon energy the cross
section for the photoemission is quite low, we used a pass energy of 5 eV in the 
single pass electron energy analyser, in order to have a good signal 
to noise ratio, which lowered the overall energy resolution to 120 meV.

\section{Theoretical} 

For the density functional theory (DFT) calculations of the SCLSs we
employ the generalized gradient approximation (GGA) of the exchange-correlation
functional \cite{perdew96}, using the full-potential linear augmented
plane wave method (FP-LAPW) \cite{blaha99,kohler96,petersen00} for
solving the Kohn-Sham equation. The Ru(0001) surface is modeled using
a six layer slab, and O is adsorbed on both sides to preserve mirror
symmetry. A vacuum region corresponding to five Ru interlayer spacings
($\approx$11{\AA}) was employed to decouple the surfaces of consecutive
slabs in the supercell approach. Within a $(2 \times 2)$ surface unit
cell, the positions of all O adatoms and Ru atoms in the outer two substrate
layers were fully relaxed for all coverages considered. The resulting
adsorption geometries are in very good agreement with existing LEED data
\cite{lindroos89,pfnuer89,gsell98,stampfl96a}, as well as with earlier
DFT pseudo-potential calculations \cite{stampfl96b}.

The FP-LAPW basis set is taken as follows: $R_{\rm{MT}}^{\rm{Ru}}=$2.3 bohr,
$R_{\rm{MT}}^{\rm{O}}=$1.3 bohr, wave function expansion inside the muffin
tins up to $l_{\rm{max}}^{\rm{wf}} = 12$, potential expansion up to
$l_{\rm{max}}^{\rm{pot}} = 4$. The Brillouin zone integration for the
$(1 \times 1)$ cells was performed using a $(12 \times 12 \times 1)$
Monkhorst-Pack grid with 19 {\bf k}-points in the irreducible part. For the
larger surface cells, the grid was reduced accordingly, in order to obtain
the same sampling of the reciprocal space. The energy cutoff for the plane
wave representation in the interstitial region between the muffin tin spheres
was 17 Ry for the wave functions and 169 Ry for the potential.

The SCLS, $\Delta_{\rm{SCLS}}$, is defined as the difference in energy which is
needed to remove a core electron either from a surface or from a bulk atom:

\begin{eqnarray}
\Delta_{\rm{SCLS}} &=&
\left[ E^{\rm{surface}}(n_c - 1) - E^{\rm{surface}}(n_c) \right] \; - \; \nonumber \\
&-& \left[ E^{\rm{bulk}}(n_c - 1)    - E^{\rm{bulk}}(n_c) \right] \quad,
\end{eqnarray}

\noindent
where $E^{\rm{surface/bulk}}(n_c)$ is the total energy of the system
considered as a function of the number of electrons, $n_c$, in a particular
core level, $c$, of a surface or bulk atom respectively \cite{spanjaard85}.
Within the initial state approximation, $\Delta_{\rm{SCLS}}^{\rm{initial}}$
is given by

\begin{equation}
\Delta_{\rm{SCLS}}^{\rm{initial}} \;\approx\;
- \left[ \; \epsilon_c^{\rm{surface}}(n_c) \; - \;
            \epsilon_c^{\rm{bulk}}(n_c) \; \right] \quad.
\end{equation}

\noindent
Here, $\epsilon_c^{\rm{surface}}$ and $\epsilon_c^{\rm{bulk}}$ are
the Kohn-Sham eigenvalues of the particular core state, $c$, so that
in this approximation the SCLS is simply due to the variation of the
orbital eigenenergies before the excitation of the core electron.
A full calculation of the ionization energy, which includes the screening
contributions from the valence electrons in response to the created
core hole, can be achieved by calculating the total energy of an
impurity with a core hole in the selected core state. The SCLS is
then the difference of two total energies, with the impurity once
located at the surface and once inside the bulk \cite{johansson80}.
To a good approximation, this difference can also be obtained
via the Slater-Janak transition-state approach of evaluating total
energy differences \cite{janak78}. Using the mean value
theorem of integration,

\begin{eqnarray}
E(n_c-1) - E(n_c) &=& \int_{n_c}^{n_c-1} \frac{\partial E(n')}{\partial n'} dn'
\; \approx \nonumber \\
&\approx&  - \epsilon_c(n_c - 1/2) \quad,
\end{eqnarray}

\noindent
eq. (1) can be cast into the form of eq. (2), yet this time with a core
level occupation of $n_c - 1/2$. Note that this latter
approach, from which we derive what we will henceforth call the total SCLS,
takes both initial and final state effects (in the spectroscopic sense)
into account, so that the results can be compared with the experimental
values.

Whereas initial state SCLSs can directly be obtained from our normal
all-electron scheme, the total SCLSs require a self-consistent impurity
calculation, where one atom is ionized by removing half an electron from
the considered core level. We used $(2 \times 2)$ supercells to surround
each such atom with neighbours possessing the normal core configuration
and kept the fully relaxed ground state geometry fixed. In order to describe
an electronically fully relaxed final state, suitable for a system like
Ru with a Fermi reservoir of electrons, overall charge neutrality must be
imposed, i.e. half an electron was added at the Fermi level. 

Initial state and full calculations for the $3d$ SCLSs were done for
each inequivalent Ru atom in the outermost two substrate layers at all
experimentally described coverages. The bulk core level position,
$\epsilon_c^{\rm{bulk}}$, was calculated using a ten layer bulk slab
inside the same supercell as used for the surface calculations,
i.e. the previous vacuum region was simply replaced by additional
Ru layers. With this procedure an identical sampling of reciprocal
space was achieved for both surface and bulk calculations. Having
evaluated both the initial state and the total SCLS allows to extract
the screening contribution, which is not accessible from the
experimental data.

\section{Results}

\subsection{SCLS analysis}

\begin{figure}
       \epsfxsize=0.47\textwidth \centerline{\epsfbox{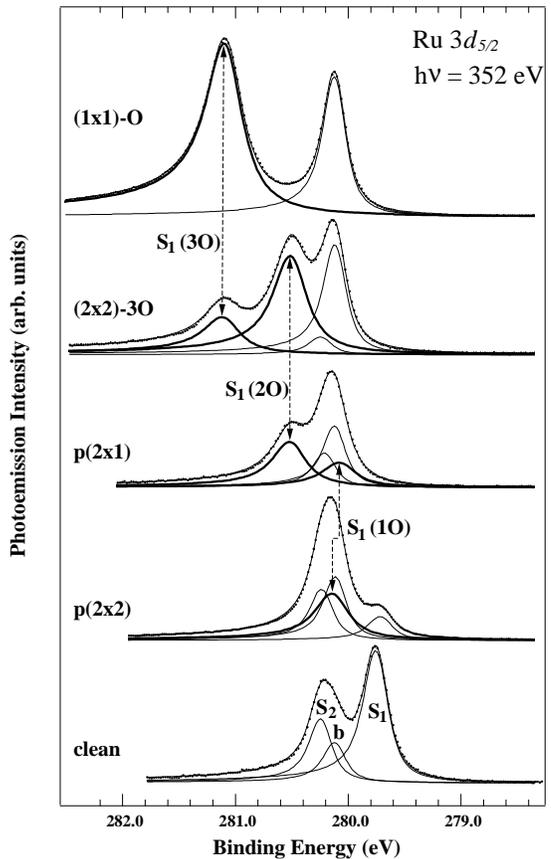}}
        \caption{Ru $3d$$_{5/2}$ core level spectra for the clean surface 
        and the four oxygen structures. The dots represent the experimental
        results, while the line in between is the result of the fit. The
        spectra were measured at a temperature lower than 130\ K. The
        components used in the fit are added in the figure. The curves with
        the thin line denote the ``clean'' components ($S_1$, $S_2$), while
        the thicker lines are the oxygen related components, $S_1$(1O),
        $S_1$(2O), and $S_1$(3O), corresponding to first layer Ru atoms
        bonded to one, two and three oxygen atoms respectively (cf. Fig. 1).
        The dashed lines with arrows denote the presence of each of these
        components in two different structures.
        \label{fig:2}}
\end{figure}

In Fig. 2 the SCLS spectra measured at 352 eV are shown, together with the
fits and the various components. The data were fitted using Doniach-Sunjic
functions convoluted with Gaussian broadening \cite{doniach70}. The
background was assumed to be linear. In order to get physically
meaningful results from the fits it was necessary to put constraints
on some parameters of the fitting function as many components have significant
overlap. The three spectra at different photon energy of a certain structure
(except for the $p(2\times 2)$) were hence fitted together with identical
parameters, leaving free only the intensities of the core level components.
In this way the line shape parameters found (Gaussian and Lorentzian width,
as well as the asymmetry parameter) are more reliable. Two strategies were
then employed to assign the various peaks to the differently coordinated
Ru atoms in the surface:

(i) Strategy (i) is an independent experimental assignment, which uses
only the structural knowledge of the various O phases as described above.
Recurrently working down in coverage starting from the $(1 \times 1)$-O/Ru(0001)
phase, all peaks can thus uniquely be identified with the notable exception of
the assignment of the S1 and S2 peak of the clean surface. The latter
determination was achieved by supplementary photoelectron diffraction
experiments, which will be described in the next subsection.

(ii) Strategy (ii) relies partially on information from our theoretical
calculations, the main difference being the inclusion of (small) non-zero
shifts of the $S_2(\rm{1O})$ peak, which was neglected in strategy (i) to
avoid overfitting. As will be discussed in section IV C, approach (ii) improves
the quantitative agreement between theory and experiment, yet we argue
that approach (i) was also important in order to assure that both,
measurement and calculation, lead independently to the same conclusions.

Details of these two fitting procedures are described in the appendix,
while the SCLS values are collected in Table I.
The error bars shown in the table were estimated from the quality
of the fits when changing the SCLS in this energy range. Therefore,
possible errors related to the oxygen coverage are not included in the table.

\begin{table}
\caption{\label{tableI}
Measured SCLSs of the Ru $3d_{5/2}$ level at all coverages in meV.
Positive shifts reflect a more strongly bound core level at the surface
compared to the bulk. The nomenclature for the different substrate
atoms ($S_1$, $S_2$ etc.) follows that of Fig. 1. In 
strategy (i) the value of the $S_2$(1O) was set to 0 for all the 
structures, while only for the $(1\times 1)$-O its value was 
obtained by fitting strategy (ii).}

\begin{tabular}{l | r | r}
                            & strategy (i) & strategy (ii)  \\
                            &              &             \\ \hline
clean, $S_1$                &  $-366\pm 10$& $-360\pm 10$\\
clean, $S_2$                &  $+125\pm 10$& $+127\pm 10$\\
$p(2\times 2)$, $S_1$       &  $-400\pm 20$&             \\
$p(2\times 2)$, $S_1$(1O)   &  $+20\pm 30$ &             \\
$p(2\times 2)$, $S_2$       &  $+120\pm 30$&             \\
$p(2\times 1)$, $S_1$(1O)   &  $-50\pm 30$ &             \\
$p(2\times 1)$, $S_1$(2O)   &  $+390\pm 10$&             \\
$p(2\times 1)$, $S_2$       &  $+88\pm 30$ &             \\
$(2\times 2)$-3O, $S_1$(2O) &  $+387\pm 20$&             \\
$(2\times 2)$-3O, $S_1$(3O) &  $+980\pm 10$&             \\
$(2\times 2)$-3O, $S_2$     &  $+127\pm 30$&             \\
$(1\times 1)$-O, $S_1$(3O)  &  $+960\pm 10$& $+920\pm 10$\\
$(1\times 1)$-O, $S_2$(1O)  &  $0$         &  $-60\pm 10$\\
\end{tabular}
\end{table}

\subsection{SCLS assignment}

\begin{figure}
       \epsfxsize=0.47\textwidth \centerline{\epsfbox{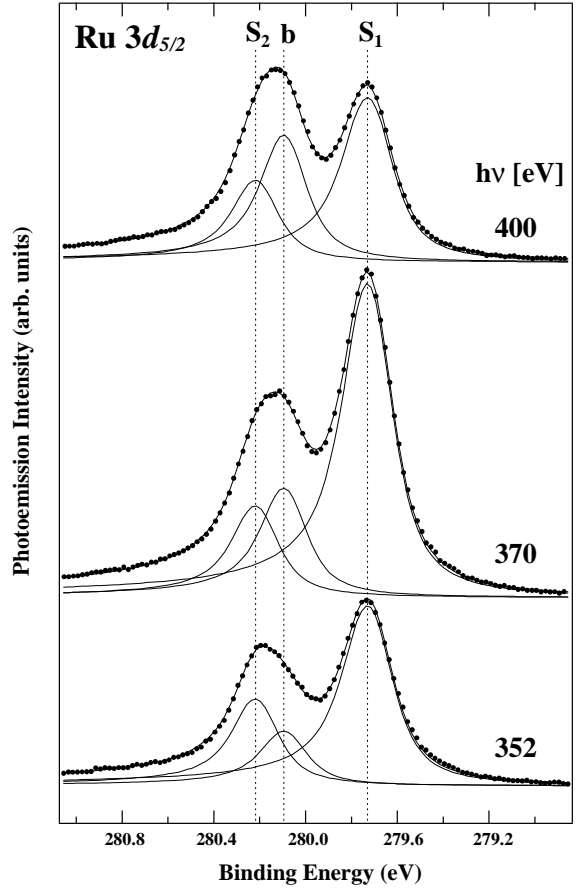}}
        \caption{SCLS spectra of the clean surface measured at different 
        photon energies. The result of the fit is added in the figure as
        a line crossing the experimental points represented by dots. The
        three components $b$, $S_1$, and $S_2$ are also added as solid
        lines. The energy range used to fit the data is wider than what
        is shown in the figure (see text). 
        \label{fig:3}}
\end{figure} 

As already mentioned, the assignment of the oxygen related SCLSs following
strategy (i) was implicit in the measurements, whereas that of the clean
surface still needs to be proved. In Fig. 3 the three SCLS spectra of the
clean surface, measured at the three photon energies given, are shown together 
with the fits. The spectra have been normalised at the low binding 
energy side. They have been measured and fitted between 277.9 eV and 
281.8 eV in a wider range than shown in the figure. Among the three peaks
present, the only one which can be unambiguously assigned is peak $b$, which
belongs to the bulk. This results from the analysis of the SCLSs of the
$(1\times 1)$-O and is also supported by the fact that when saturating the
surface with CO or other adsorbates, the only peak which remained unchanged
was peak $b$.

From a simple inspection of the data it is possible to see that peak $b$
increases at higher photon energy, consistent with a simple mean free path
picture. The peak at lower binding energy, $S_1$, has maximum intensity at
370 eV and the component at higher binding energy, $S_2$, is more or less
constant. From these data it would not be possible to disentangle the various
components accounting only for inelastic scattering effects. In fact, the
strong modulation of the lower binding energy peak, which will be assigned
to the top layer as we show in the following, is mainly due to interference 
effects, i.e. to photoelectron diffraction, and not to inelastic damping. 
Therefore we used these interference effects to find the assignment for 
the clean surface.

The approach relies on the fact that at photoelectron kinetic energies higher
than $\approx$400 eV the conditions for strong forward scattering are
fullfilled when an atom of the first layer lies in the line between a second layer
emitter atom and the electron energy analyser (cf. Fig. 1, top right panel) 
\cite{woodruff94}. Therefore, changing the azimuthal
angle $\phi$ at an appropriate polar angle $\theta$ (for the clean Ru(0001)
$\theta$=$36^{\circ}$) one should see that the  photo-emission intensity of the
second layer is strongly modulated due to the forward scattering with the first
layer, while the intensity of the peak due to the latter atoms stays almost
constant since no scatterers are present between the emitter in the first layer
and the analyser \cite{lizzit98}. The problem, which arises in this
experiment, is that at such
a high kinetic energy and low emission angle, the intensity of the photoemission
from the first layers will decrease appreciably with respect to that from the bulk.
This will affect much more the $S_2$ peak, which is very close to the bulk peak,
thus becoming almost undetectable. 

\begin{figure}
       \epsfxsize=0.47\textwidth \centerline{\epsfbox{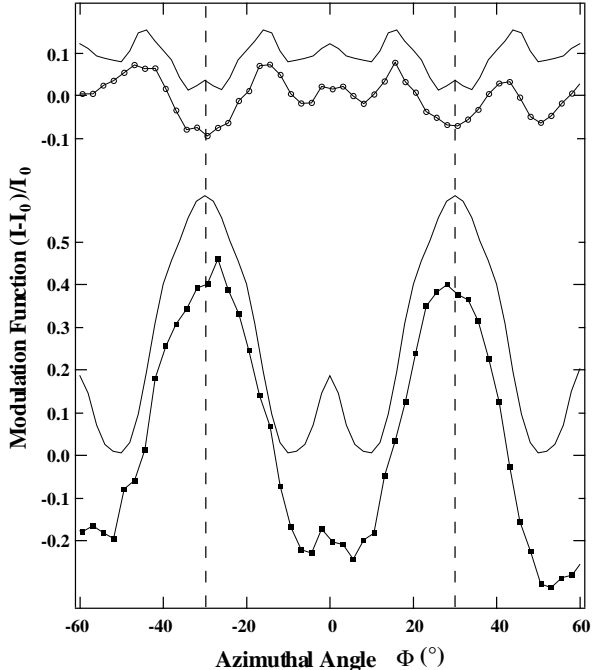}}
        \caption{Angular dependence of the modulation function of
        the $S_1$ (open circles) and $S_2$ (filled squares) components
        shown in Fig. 3. The $S_2$ component shows a clear enhancement
        of the intensity in the forward scattering directions, denoted
        by the dashed lines in the figure. The solid lines represent 
        the results of multiple scattering simulations. These two curves 
        have been shifted with respect to each other for display purposes.
        \label{fig:4}}
\end{figure} 

In order to overcome this problem, we performed preliminary multiple 
scattering simulations of the first and second layer photoemission 
intensity. We used the MSCD package developed by Chen and Van Hove \cite{vanhove} 
which uses multiple scattering theory and the Rehr-Albers separable
representation of spherical waves propagators \cite{rehralbers}. As input 
to the program we used the structural parameters obtained from a 
previous LEED I/V experiment \cite{over92}. Moreover, since the Ru(0001) surface 
is composed by domains rotated by $120^{\circ}$ to each 
other, we summed the photoemission intensity over these domains.
At the end we calculated the modulation function defined as 
$(I(\phi)-I_0)/I_0$, where $I(\phi)$ is the photoemission intensity,
while $I_0$ is its average value.
From these calculations we found the best conditions to perform the 
photoelectron diffraction experiment.  In particular, we realized that when
performing an azimuthal scan at $\theta$=$40^{\circ}$ at a kinetic energy 
of 220 eV, not only the first layer intensity shows pronounced 
modulations due to the backscattering, but furthermore these are 
in antiphase with those of the second layer emission in which the 
characteristic forward scattering peaks are present at 
$\phi$=$\pm 30^{\circ}$ respect to the [1$\bar2$10] direction.  
The photoelectron diffraction experimental results together with the 
multiple scattering simulations are shown in Fig. 4.
The agreement between experiment and simulation is very good, hence 
giving a clear answer to the question we addressed: $S_1$ belongs to 
the first layer atoms, while $S_2$ to those of the second layer.

\subsection{Comparison with theory}

Having achieved an unambiguous assignment of all experimentally
detected peaks, the next step is to compare these results with
the calculated SCLSs. As our intention is to make use of the possibility
to decompose the latter shifts into initial and final state
contributions, the agreement between theory and experiment should not
only be on a qualitative or semi-quantitative level, but should convincingly
make clear that there are no inconsistencies whatsoever between both
data sets.

In order to perform such a comparison, we first address the accuracy
of the DFT calculations. Possible numerical errors can arise due
to the use of a finite basis set, as well as due to the finite size of
slab and vacuum region in the supercell approach. To assess the effect
on the derived SCLS values, we sequentially increased the corresponding
values and monitored the SCLSs of both first and second layer atoms of the
clean and $(1 \times 1)$-O covered surface, which form the lower and upper
bound of the coverage sequence considered. We checked the convergence of the
basis set by increasing the plane wave cutoff in the interstitial from
17 Ry to 23 Ry, as well as using denser {\bf k}-meshes up to a
$(18 \times 18 \times 1)$ grid with 37 {\bf k}-points in the irreducible wedge.
In both cases the SCLS changes were within $\pm 10$ meV. As the SCLSs result
from a difference between a surface and a bulk quantity, the obvious point
here is to use exactly the same basis set in both calculations, which then
leads to a good cancelation of errors and thus makes the SCLS value itself
less sensitive to the finite FP-LAPW basis set used.

The main source of error due to the supercell approach stems from
the use of slabs of finite thickness. Test calculations performed
with ten layer slabs revealed changes in the SCLSs up to $\pm 20$ meV,
particularly in the second layer shifts. As the changes in the calculated
work function were of the same order, we assign these differences to
slight variations of the Kohn-Sham potential due to a quantum size
effect in the finite slab. On the other hand, further
increasing the vacuum region did not have any influence on the SCLS
values ($< \pm 5$ meV). Summarizing both the errors due to the
basis set and the supercell approach, we hence give a conservative
estimate of the numerical accuracy of $\pm 30$ meV, which is of the same
order as the experimental error, thus justifying the chosen setup.

However, this error estimate does not comprise possible errors due to
general deficiencies of the approach, i.e. due to the selected
exchange-correlation potential or the use of the transition-state concept
to evaluate the total shifts. To this end, we also calculated the SCLSs 
for both $(1 \times 1)$ phases using the local density approximation (LDA)
for the exchange-correlation functional \cite{perdew92}. We found the
$S_1$ and $S_2$ of the clean surface, as well as the $S_2(\rm{1O})$ of
the $(1 \times 1)$-O phase to lie within $\pm 10$ meV of the values
obtained with the GGA. On the other hand, the SCLS of the threefold O
coordinated first layer atom $S_1(\mbox{3O})$ changed by 101 meV,
significantly worsening the agreement with the experimental value.
We attribute this finding to an improved description within the GGA,
which -- as deduced from the remarkable agreement between
experiment and theory reported below -- seems to allow a highly
accurate determination of the quantity of interest to our study.

\begin{figure}
       \epsfxsize=0.47\textwidth \centerline{\epsfbox{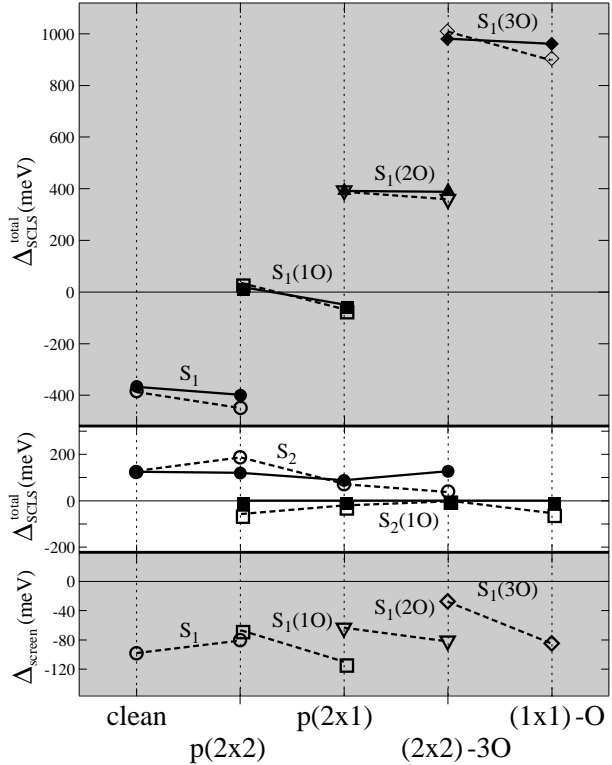}}
        \caption{Comparison of the calculated SCLSs (open symbols) with 
        the experimental results (filled symbols) obtained by fitting strategy
        (i). The top panel represents the SCLSs of the first substrate layer Ru atoms,
        while the middle panel displays the SCLSs of the second layer
        ones. The bottom panel displays the screening contribution to
        the total first layer shifts.
        \label{fig:5}}
\end{figure} 

Figure 5 shows a comparison between the calculated and the measured
SCLSs. It is immediately obvious that almost all theoretical and experimental
shifts fall within their mutually assigned error bars, reflecting the
consistency between both data sets we aim at. While still showing a
good semi-quantitative agreement, only the following shifts do not
meet this requirement: $S_2$ and $S_2$(1O) in the $p(2\times 2)$,
$S_2$ in the $(2\times 2)$-3O, as well as $S_1$(3O) and $S_2$(1O) in the
$(1\times 1)$-O. The disagreement in the $S_2$(1O) shifts is not surprising,
as this component was neglected in the original experimental data
analysis ( strategy (i), cf. section IV A ) in order to avoid overfitting.
After the theoretical calculations had predicted non-vanishing $S_2$(1O)
shifts particularly for the $p(2\times 2)$ and the $(1\times 1)$-O phases,
the experimental data set was reanalyzed including this component
( strategy (ii) ). This was unambiguously possible in the case of the
$(1\times 1)$-O phase with its clearly separate bulk and surface peaks.
The resulting value of $S_2(\rm{1O}) = -60 \pm 10$meV agrees perfectly 
with the theoretical $S_2(\rm{1O}) = -53 \pm 30$meV, bringing now also
the calculated and measured $S_1$(3O) peak into consistency (theory:
$ +899 \pm 30$meV, exp: $920 \pm 10$meV). Unfortunately, the crowding of
peaks around the bulk peak in the $p(2\times 2)$ phase did not allow to
add yet another component to the fitting procedure. Hence, we were not
able to resolve the small discrepancy for the $S_2$(1O) peak in this phase.

This leaves only the $S_2$ components in the $p(2\times 2)$ and in the
$(2\times 2)$-3O. As just discussed, the experimentally derived value for
the $p(2\times 2)$ could be affected by neglecting the $S_2$(1O)
peak in the fitting procedure. Additionally, this structure was measured
only at 352 eV, and furthermore probably the error bar of the measured SCLS
is bigger due to the presence of many peaks in a very small energy
range. This can then certainly account for the small difference of
67 meV between calculated and measured shift. Yet, these reasons do not
apply in the case of the $(2\times 2)$-3O, where theory predicts a vanishing
$S_2$(1O) shift and which was measured at three photon energies. Here,
however the weight of the $S_2$ component is quite small compared to the
others, thus increasing the error in the experimental determination of
its position. Under these circumstances we do not consider the small
difference of 88 meV between theoretical and experimental shift to reflect
a significant inconsistency. In conclusion, we hence find both data sets
to be fully compatible with each other.

\section{Analysis}

\begin{table} 
\caption{\label{tableII} 
Calculated Ru $3d$ SCLSs for the first layer atoms at various coverages. 
Shown are the total shifts, as well as their decomposition into screening 
and initial state parts: 
$\Delta_{\rm{SCLS}}^{\rm{total}} = \Delta_{\rm{screen}} + 
 \Delta_{\rm{SCLS}}^{\rm{initial}}$. The rightmost column contains the 
initial state shifts as obtained for Ru bulk truncated geometries. Units 
are meV.}  

\begin{tabular}{l | r | rr || r} 
                           & Total        & Screening  & Initial        & Initial          \\ 
                           &              &            & (relaxed)      & (bulk-trunc.)    \\ \hline 
clean, $S_1$               &  -383        &    -98     &     -285       &  -338            \\ 
$p(2\times 2)$, $S_1$      &  -448        &    -80     &     -368       &  -407            \\ 
$p(2\times 2)$, $S_1$(1O)  &   +36        &    -65     &     +101       &   +42            \\ 
$p(2\times 1)$, $S_1$(1O)  &   -67        &   -111     &      +44       &   -12            \\ 
$p(2\times 1)$, $S_1$(2O)  &  +395        &    -62     &     +457       &  +454            \\ 
$(2\times 2)$-3O, $S_1$(2O) &  +362        &    -80     &     +442       &  +476            \\ 
$(2\times 2)$-3O, $S_1$(3O) & +1010        &    -27     &    +1037       & +1088            \\ 
 $(1\times 1)$-O, $S_1$(3O) &  +899        &    -85     &     +984       & +1072            \\ 
\end{tabular} 
\end{table} 
 
\subsection{Screening effects}

While a main idea behind the study of SCLSs is to gain an understanding
of the electronic and structural environment of atoms at the unperturbed
surface, i.e. before the core excitation, the measured shifts comprise
an additional component, which is due to the different screening
capabilities of the core-ionized system at the surface and in the
bulk \cite{spanjaard85}. In fact, this screening capability is closely
related to the electronic hardness and the surface chemical activity
(see e.g. Stampfl et al. \cite{stampfl01} and references therein); thus,
also this information is of significant interest. Fortunately, calculations
as applied in this work provide the possibility to separate the total
(measured) shifts into the initial state and the additional final state
(i.e. screening) contributions. Table II lists these components for all
first layer atoms at the coverages considered. We see that the magnitude
of the screening correction is rather small compared to the overall trend in
the initial state shifts. Although it leads to an enhanced difference in the
total shifts of equally coordinated Ru atoms particularly in the case of the
$S_1(\rm{1O})$ and $S_1(\rm{3O})$ atoms, it still does not overshadow the
clear dependence on the number of direct O neighbours, cf. Fig. 5. However,
this does not imply that it could be neglected, as only the full shifts lead
to the good agreement with the experimental data described above: The initial
state shifts alone fall far out of the experimental error bars. Note
that especially in the case of the small total shifts corresponding to
singly O-coordinated Ru surface atoms, the screening contribution is even
larger in magnitude than the initial state shift.

\begin{table} 
\caption{\label{tableIII} 
Calculated Ru $3d$ SCLSs for the second layer atoms at various coverages. 
Shown are the total shifts, as well as their decomposition into screening 
and initial state parts: 
$\Delta_{\rm{SCLS}}^{\rm{total}} = \Delta_{\rm{screen}} + 
 \Delta_{\rm{SCLS}}^{\rm{initial}}$. Units are meV.} 
 
\begin{tabular}{l | r | rr } 
                           & Total        & Screening  & Initial  \\ \hline 
clean, $S_2$               &  +124        &    -72     &     +196 \\  
$p(2\times 2)$, $S_2$      &  +187        &    -19     &     +206 \\  
$p(2\times 2)$, $S_2$(1O)  &   -57        &    -82     &      +25 \\  
$p(2\times 1)$, $S_2$      &   +72        &    -34     &     +106 \\  
$p(2\times 1)$, $S_2$(1O)   &   -21        &    -96     &      +75 \\  
$(2\times 2)$-3O, $S_2$     &   +39        &    -44     &      +83 \\  
$(2\times 2)$-3O, $S_2$(1O)  &    +3        &    -35     &      +38 \\  
 $(1\times 1)$-O, $S_2$(1O)  &   -53        &    -83     &      +30 \\  
\end{tabular} 
\end{table} 
 
This is even more so for the small total shifts connected to second layer Ru
atoms ( $S_2$ and $S_2$(1O) ). Here, the screening correction is of the same
order of magnitude as the initial state shift itself ($\approx 100$ meV),
and similar to the trend found for the first layer atoms always negative in
sign (cf. Table III). As all initial state $S_2$ and $S_2$(1O) are found to
be positive, frequent sign changes are hence introduced by the screening
contribution. Consequently, in the measurement the second layer shifts can
lead to small peaks in close vicinity on {\em either side} of the bulk peak,
which will be hard to resolve experimentally. As is apparent
from the two fitting procedures employed in the present experimental analysis
(cf. section IV A), this can then indirectly also influence the assessment of
the larger first layer shifts. Given that the latter are typically the ones
of primary interest, special care with respect to this point should therefore
be exerted in the experimental data analysis. 

\begin{figure}
       \epsfxsize=0.45\textwidth \centerline{\epsfbox{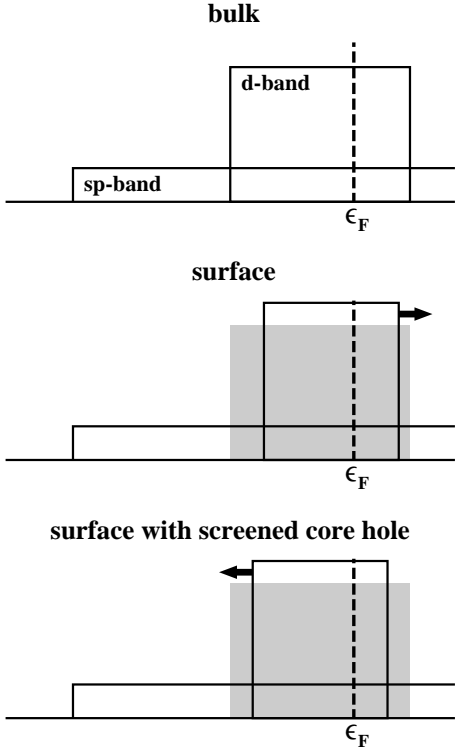}}
        \caption{Schematic DOS in the rectangular $d$-band model (for the
        case of a more than half full $d$-band). At the surface the
        $d$-band is narrowed and shifted up in energy to maintain local
        charge neutrality. Upon core excitation the $d$-DOS shifts to lower
        energies and a valence electron from the Fermi reservoir restores
        local charge neutrality by filling up formerly unoccupied $d$-states.
        The enhancement of the surface $d$-DOS at and above the Fermi level
        leads to a more efficient screening at the surface and hence to a
        negative screening contribution to the total SCLS. Note that in the
        case of a less than half full $d$-band the $d$-DOS is shifted down
        in energy due to the narrowing and hence a negative initial state
        contribution to the SCLS results. However, the enhancement of the
        $d$-DOS at and above the Fermi level nevertheless leads to a negative
        screening contribution.
        \label{fig:6}}
\end{figure} 

Methfessel and coworkers have shown that final state effects at clean, true
transition metal surfaces are largely due to intra-atomic $d$-electron
screening \cite{andersen94,methfessel93,methfessel95}. Upon core excitation,
the $d$-DOS shift to lower energies causes a valence electron from the Fermi
reservoir to restore local charge neutrality by filling up formerly
unoccupied $d$-states. Due to the lowered coordination at the surface, the
local density of $d$-states ($d$-DOS) is narrower in energy compared to the
$d$-DOS of a bulk atom. Because the total number of states in a band is
conserved, already in the simplest rectangular $d$-band model with
a constant $d$-DOS \cite{friedel69} one would then expect the $d$-DOS
value at and above the
Fermi level to be enhanced compared to the bulk situation. This is
schematically shown in Fig. 6. In turn, this enhancement implies
that the core hole be more efficiently screened at the surface, which
in our present sign convention leads to a negative screening correction. 
In Fig. 7 we show the real self-consistent $4d$-DOS, calculated inside the
muffin tin spheres \cite{remark1} for the two limiting phases of the
considered coverage range, i.e. the clean and the $(1 \times 1)$-O surface.
Compared to the bulk situation, we indeed find the clean surface $d$-DOS
to be narrowed in energy and in the energy range at and above the Fermi level
it is strongly enhanced. Despite the widening of the $d$-band caused by the O
adsorption (see below), this enhancement prevails also for all O covered
surfaces, exemplified in Fig. 7 with the $(1 \times 1)$-O phase. Consequently,
negative screening contributions are found throughout the whole coverage
sequence.

\begin{figure}
       \epsfxsize=0.47\textwidth \centerline{\epsfbox{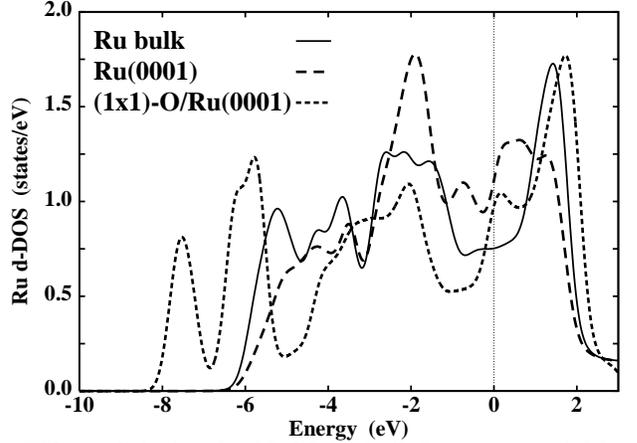}}
        \caption{Calculated $4d$-DOS for bulk Ru atoms (solid
        line) and for first layer Ru(0001) atoms of the clean
        (dashed line) and $(1 \times 1)$-O covered surface
        (dotted line). The energy zero is at the Fermi level.
        \label{fig:7}}        
\end{figure} 

It is interesting to compare this situation to the work for
O adlayers on Rh(111)\cite{ganduglia00}. There, a sign change in the
screening contribution was found, with the lower coverage surfaces screening
again better than the bulk, but the higher O-covered surfaces screening
worse (cf. Fig. 8). This is connected to the fact that in Rh, which is
situated just right of Ru in the periodic system, the Fermi level
is located at a different position in the $4d$-band. Above that position,
the $d$-DOS is lowered so strongly upon O adsorption that it eventually
falls below the value of the bulk $d$-DOS and thus induces the sign change
in the screening correction. In Ru on the other hand, this lowering never
reaches the bulk $d$-DOS, so that the screening remains negative in
sign throughout (cf. Fig. 8).

\subsection{Initial state shifts}

\begin{figure} 
       \epsfxsize=0.47\textwidth \centerline{\epsfbox{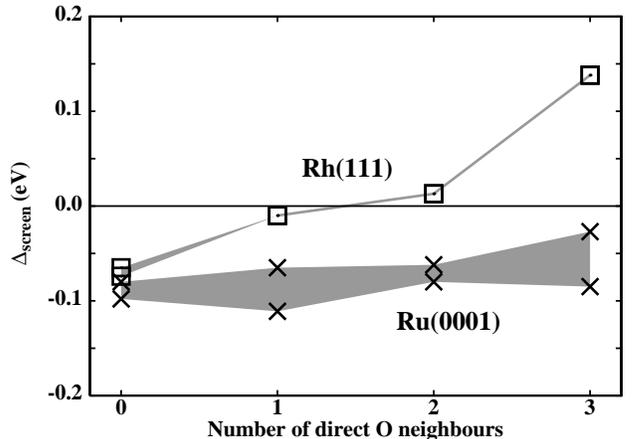}} 
        \caption{Comparison of the screening contribution, 
        $\Delta_{\rm screen}$, for O/Ru(0001) (crosses) and 
        O/Rh(111) (boxes) as a function of the number of 
        directly coordinated O atoms. The shaded area is 
        drawn to guide the eye. The O/Rh(111) data is taken 
        from \cite{ganduglia00}. 
        \label{fig:8}} 
\end{figure}  
 
Having subtracted off the final state effect from the total SCLSs,
we are now in a position to discuss the initial state contribution, i.e.
the change in the local (near nucleus) electrostatic field (see below).
For clean transition metals, these shifts are well understood in terms of
the narrowing of the surface valence $d$-band due to the lowered coordination
\cite{spanjaard85}. In order to maintain local charge neutrality, the center
of a less (more) than half full $d$-band moves down- (up-) wards in energy,
which goes hand in hand with an attractive (repulsive) contribution to the
Kohn-Sham potential (cf. Fig. 6). This potential change acts on the core
electrons as well and induces a positive SCLS for the early and a negative
SCLS for the late transition metals. This trend involving a sign change across
the series is well confirmed by a number of experimental and theoretical studies
\cite{spanjaard85,andersen94,methfessel93,alden93}, into which the here
derived negative $\Delta_{\rm{SCLS}}^{\rm{initial}}$ for clean Ru(0001) fits nicely.

Upon O adsorption, the O $2p$ level interacts with the localized Ru $4d$
states, causing the formation of bonding and antibonding states close to
the lower and upper edge of the valence $4d$-band respectively (cf. Fig.
7) \cite{scheffler00}. The ensuing increased width of the valence band
requires then again an adjustment of the center of gravity of the band in
order to maintain local charge neutrality. In the following we will show
that this adjustment moves the band downwards in energy and the corresponding
attractive contribution to the Kohn-Sham potential is reflected in more and
more positive SCLSs with increasing O coverage. Further, as the width is
connected to the formation of bonds, which obviously scale with
the number of directly bound O atoms, similar SCLSs result for equally O
coordinated Ru atoms.

\begin{table}
\caption{\label{tableIV}
Shift of the center of gravity, $\Delta C_{4d}$ in meV, and relative change
in the width, $\Delta W$, of the Ru valence $4d$-band for all first layer
atoms at the coverages considered with respect to the bulk situation.
Additionally shown in the middle column is the shift of the center of gravity
resulting from a simple rectangular $d$-band model as described in the text.}

\begin{tabular}{l | rr | r}
                           &  $\Delta C_{4d}$ & $\Delta \tilde{C}_{4d}$ &  $\Delta W$ \\
                           &                  & (model)         &             \\ \hline
clean, $S_1$               &   -200           &   -200          &    -12\%    \\
$p(2\times 2)$, $S_1$      &   -180           &   -180          &    -11\%    \\
$p(2\times 2)$, $S_1$(1O)  &      0           &    +30          &     +2\%    \\
$p(2\times 1)$, $S_1$(1O)  &    -20           &    +50          &     +3\%    \\
$p(2\times 1)$, $S_1$(2O)  &   +140           &   +220          &    +13\%    \\
$(2\times 2)$-3O, $S_1$(2O) &   +160           &   +250          &    +15\%    \\
$(2\times 2)$-3O, $S_1$(3O) &   +480           &   +480          &    +29\%    \\
 $(1\times 1)$-O, $S_1$(3O) &   +410           &   +480          &    +29\%    \\
\end{tabular}
\end{table}

In order to quantify this trend, we have evaluated the first and second
moment of the valence $4d$-band for each first layer atom at the coverages
considered. The $p$th moment of the DOS, $N(\epsilon)$, is defined as
\cite{pettifor95},

\begin{equation}
\mu_p \; = \; \int \;N(\epsilon) \;\epsilon^p \;d\epsilon \quad,
\end{equation}

\noindent
where in our case $N(\epsilon)$ is the DOS of the Ru $4d$ states
\cite{remark1,remark2}. $\mu_0$ gives the total number of states in the band
and $\mu_1/\mu_0 = \epsilon_{4d}$ its center of gravity. Having obtained
these moments for all coverages and for the bulk, allows us then to
calculate the shift of the band,
$\Delta C_{4d} = \epsilon_{4d}^{\rm{bulk}} - \epsilon_{4d}^{\rm{surf}}$,
with respect to the bulk situation. The second moment, $\mu_2/\mu_0$,
is proportional to the mean square width, $W^2$, of the band which we
again translate into relative width changes, 
$\Delta W = W^{\rm{surf}}/W^{\rm{bulk}} - 1$, with
respect to the bulk situation. As shown in Table IV, the not O coordinated
$S_1$ atoms possess a band which is 12\% narrower than the bulk one,
and correspondingly it is shifted by $\approx 0.2$ eV to higher energies
(cf. Fig. 7). On the other hand, the threefold O coordinated
$S_1(\mbox{3O})$ atoms have a band, which is 29\% wider than the one of
bulk Ru atoms and its center of gravity is hence shifted by $\approx 0.5$ eV
to lower energies (cf. Fig. 7).

To gain a {\em qualitative understanding} in how far the observed shift of the
center of gravity is due to the different band width, we next considered
the simplistic rectangular $d$-band model, i.e. a box of constant $d$-DOS
(cf. Fig. 6) \cite{friedel69,pettifor95}. In this model $\epsilon_{4d}$ is exactly
in the middle of the band, i.e. it is $W/2$ above the band bottom,
$\epsilon_{4d} = \epsilon_{dn} + W/2$. When this box is positioned with
respect to the Fermi level so as to achieve an ideal 70\% filling of the Ru
$4d$-band, i.e. when we impose local charge neutrality
($\epsilon_{dn} = - 0.7 W$, because the Fermi level is our energy zero),
then the width, $W$, of the box and its center of gravity, $\epsilon_{4d}$,
are related via,

\begin{equation}
\epsilon_{4d} = - \frac{2}{10} W \quad .
\end{equation}

\noindent
With the help of eq. (5), the value of the bulk center of gravity derived
from the calculated first moment determines the corresponding width and
with this the complete projection of the self-consistent bulk $d$-DOS
onto the rectangular model \cite{remark3}. After that, the differential
form of eq. (5) allows to convert the calculated relative width changes,
$\Delta W$, shown in Table IV, into relative shifts of the center of
gravity compared to the bulk situation. The resulting shifts,
$\Delta \tilde{C}_{4d}$, are given in the middle column of Table IV and match
very well the ones obtained directly from the first moment of
the real $d$-DOS. This confirms that the main driving force behind the
observed $4d$-band shift, first up in energy for the clean surface and
then lower and lower in energy upon increased O coordination, is indeed
the notion to preserve local charge neutrality upon a changing $d$-band
width.

\begin{figure}
       \epsfxsize=0.47\textwidth \centerline{\epsfbox{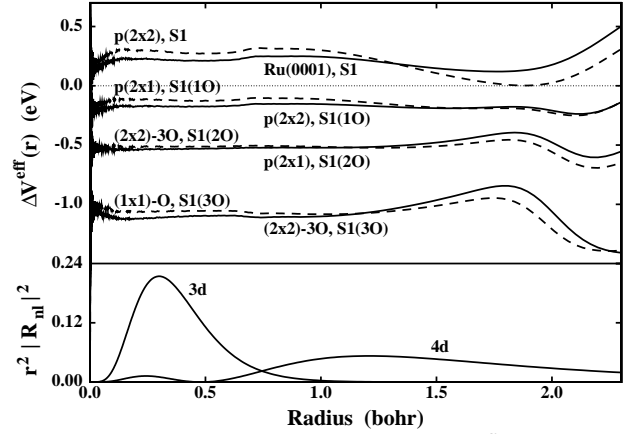}}
        \caption{Top panel: Potential shift, $\Delta V^{\rm{eff}}(r)$, inside
        all first layer Ru muffin tin spheres for the various coverages
        considered. Bottom panel: Radial part of the wavefunction,
        $ r^2 \left| R_{nl} (r) \right|^2$, for the $3d$ and $4d$ orbitals
        of bulk Ru.
        \label{fig:9}}        
\end{figure} 

The shift of the $d$-band center is accompanied by a corresponding shift
of the Kohn-Sham potential, which in turn is felt by the core electrons
and gives rise to the initial state contribution to the SCLSs. In Fig. 9
we show the spherically symmetric part of this potential shift,
$\Delta V^{\rm{eff}}(r)$,

\begin{equation}
\Delta V^{\rm{eff}}(r) \;=\; V^{\rm{eff}}_{\rm{surf}}(r) - 
                             V^{\rm{eff}}_{\rm{bulk}}(r) \quad,
\end{equation}

\noindent
as a function of the radial distance, $r = | \bf{r} - \bf{R} |$,
from the nucleus at $\bf{R}$. The shift is primarily related to the
number of directly coordinated O atoms, starts with positive
shifts (more repulsive potential) for the $S_1$ type atoms and turns into
more and more negative shifts for the $S_1(\rm{1O})$, $S_1(\rm{2O})$ and
$S_1(\rm{3O})$ atoms (more attractive potential). Interestingly,
$\Delta V^{\rm{eff}}(r)$ is always almost constant up to about $\approx 1.2$
bohr away from the core. Yet, this is the region seen by the $3d$ core
electrons, as exemplified by the extension of the $3d$ radial wavefunction
for bulk Ru also plotted in Fig. 9. To first order \cite{ganduglia96},

\begin{equation}
\Delta_{\rm{SCLS}}^{\rm{initial}}(3d) \;\approx\;
 - 4\pi \int dr \; \Delta V^{\rm{eff}}(r) r^2 | R_{3d}(r) |^2
\end{equation}

\noindent
holds. Given that $\Delta V^{\rm{eff}} \approx const$ in the region of the
$3d$ orbital and the radial wavefunction is normalized,
$\Delta_{\rm{SCLS}}^{\rm{initial}}(3d) \approx - \Delta V^{\rm{eff}}$ results.
Of course, an analogous relation to eq. (7) holds also for all other deeper
lying core levels, whose $r^2 | R_{nl}(r) |^2$ are confined to an even
more localized region around the nucleus, also within the constant region
of $\Delta V^{\rm{eff}}(r)$. Hence, the different core levels all display
roughly similar shifts \cite{methfessel93}. Obviously, this is not the case
for the $4d$ valence band, which as shown in Fig. 9 has a much larger
radial extension. Hence, it reaches well into the region where
$\Delta V^{\rm{eff}}(r)$ is not constant anymore, which is mainly caused
by an increased exchange-correlation contribution in this region of lower
electron density \cite{ganduglia96}. In this region also the non-spherical
contributions to the Kohn-Sham potential become significant, so that the
magnitude of the shift of the center of gravity of the $4d$-band, $C_{4d}$,
and of $\Delta_{\rm{SCLS}}^{\rm{initial}}$ will not be similar, while
their overall trend is, as is indeed found when comparing the
values given in Table IV and Table II respectively.

Having established the relation between the measured SCLS and the local
bonding, at least to the degree as it is reflected in the valence $d$-DOS,
let us focus now on the second layer shifts. Here, only the $S_2$ type
atoms of the clean and $p(2 \times 2)$ phase display relatively large
shifts of $\approx 200$ meV, whereas the shifts of all other second layer
atoms remain very small (cf. Table III). Evaluating again the first and
second moment of the $d$-DOS for these atoms, we indeed find only the
widths for these two $S_2$ atoms increased by 5\% with respect to the bulk value
together with a corresponding shift of the $4d$-band center to lower energies,
which gives rise to their positive SCLSs. Yet, while the increased width in
the case of the first layer atoms can be explained in terms of binding to
more and more O atoms, the second layer Ru atoms always have the same number
of nearest neighbours as in the bulk. In this respect it is interesting to
notice that only the two mentioned $S_2$ atoms have first layer neighbours,
which are not yet bound to any O atom at all and which hence have somewhat
unsaturated bonds. We thus argue that these first layer atoms will most likely
reinforce their backbond to the second layer atom below, which will then
experience stronger binding than in the bulk situation. Note that this is
also reflected in the contraction of the first
layer distance with respect to the bulk, which is found only for the lower
O coverage phases \cite{lindroos89,pfnuer89}.
Judged from the width of the $d$-DOS, cf. Table IV, any Ru atom
that has established bonds to at least one O atom will not show an enhanced
backbond tendency anymore, which explains why all other second layer atoms
display a more or less bulk-like $d$-DOS width and consequently very small
SCLSs.

\section{Discussion}

The analysis of the initial state contribution just presented has shown
how the core level shifts act as a sensitive probe of the local electronic
structure around an atom, i.e. more precisely how the SCLSs are affected
by the {\em formation of bonds} between the O adsorbates and the Ru first
layer atoms. Yet, one could also hope to use the SCLSs to gain a deeper
insight into the {\em nature of the chemical bonds} between the atom
and its neighbours. Particularly in the case of adsorbates, i.e. unlike
bonding partners, it is tempting to address via the SCLSs the question of
charge transfer to or from the surface atoms, or in other words the ionic
and covalent contributions to the bonding. In the following subsection we
will first discuss our point of view on this relation between SCLSs and
charge transfer, and will thereafter apply it to interpret the bonding
situation in the O/Ru(0001) and O/Rh(111) systems.

\subsection{SCLSs and charge transfer}

In the simplest view, charge transfer off (onto) an atomic site leads to
a more attractive (repulsive) potential, thereby causing a shift in the
core level towards higher (lower) binding energy. In the case of chemisorption
of an electronegative species like oxygen, one would hence expect each time
more positive SCLSs for the higher O coordinated Ru first layer atoms, 
$S_1(\mbox{1O})$, $S_1(\mbox{2O})$, and $S_1(\mbox{3O})$ respectively,
as we indeed observe. Yet, despite this qualitatively correct trend, the
question remains whether the SCLSs could further be used to better
quantify the amount of charge actually transferred. Related to this is then
also the question whether the total adsorbate-induced shifts could really
be attributed solely to charge transfer.

Recent theories of SCLSs \cite{benesh92,bagus93,bagus99} have tried to
separate the total shift into additional factors apart from charge transfer,
namely an environmental and a configurational contribution. The former is
viewed as arising from embedding the atom into the delocalized valence
charge density of all neighbouring atoms. The ensuing overlap of these
valence orbitals onto the atomic site influences the Kohn-Sham potential
at the nucleus of the core ionized atom and thus contributes to the shift.
Note that such a contribution obviously scales with the number of neighbours,
i.e. in our case with the number of directly coordinated O atoms. The
configurational contribution, on the other hand, arises in transition
metals from the hybridization of the valence $d$-band with $sp$ states
below and above the Fermi level. The latter orbitals are much more diffuse,
i.e. the corresponding charge is on average further away from the nucleus.
Hence, a slight redistribution of electrons among these levels at the
surface can then also influence the potential. For the particular case
of ionic adsorbates on metals, also the polarization of the surface, which
tries to screen the adsorbate electric field, has been discussed
\cite{bormet94,bagus93}.

Correspondingly, the total observable shift would then be the net result of
all these (partially canceling) contributions. This argument was e.g.
employed to explain the very small negative shifts observed for alkali metal
adsorbates on W(110) in contrast to the large positive shifts caused by
O/W(110) \cite{benesh92,bagus99,riffe90}. Neglecting any other contribution
apart from charge transfer, one would in this case infer a much lower
ionicity of the electropositive alkali metals compared to the electronegative
oxygen \cite{riffe90}. Yet, this picture was contradicted by the more refined
analyses taking also environmental and configurational contributions into
account \cite{benesh92,bagus99}. In any case, although all these concepts like
charge transfer, covalent bonding or polarization are without doubt useful
for our understanding, one has also to admit that they are somehow arbitrary
(at least to a certain degree): Whether the build-up of charge between a surface
atom and an adsorbate is called covalent bonding or polarization of the metallic
charge in response to the adsorbate; or whether the overlap of valence
orbitals onto other atomic sites is already called charge transfer or not is
simply a matter of taste. In view of the analysis presented in the last
section, the very large shift of +1269 meV between the $S_1$ atoms of the
clean surface and the threefold O coordinated $S_1(\rm{3O})$ of the
$(1\times 1)$-O phase is simply the consequence of the strong interaction of
the O $2p$ orbitals with the metal $4d$ valence band, which gives rise to
bonding and anti-bonding states widening the band. That this goes hand in
hand with the sequential build-up of charge between the adsorbate and the
Ru surface atom can be seen in Fig. 9, where the surface potential shift
shows a more and more pronounced inflection in the region further than
$\approx 1.7$ bohr away from the nucleus. Interpreting this to a certain
degree as charge transfer to the O atoms would make the core-level analysis
compatible with the continuous increase of the work function upon O adsorption
\cite{stampfl96b} and with calculated charge difference density distributions.
Yet, a clear assignment of how much charge is really transferred cannot be
made on these grounds.

Coming back to the point why alkali metals show much smaller shifts, one has
also to take into account their different interaction with a transition
metal surface. The strong interaction of the O $2p$ orbitals with the Ru
$4d$-band results in a small O-Ru bondlength of $\approx 2.0${\AA}. Even the
smallest alkali metal, Li, has a bondlength of $\approx 2.7${\AA} to Ru,
reflecting a much weaker bond. The interaction with the Li $s$ orbital does
not affect the $d$-band width, and leads in turn only to very small SCLSs.
Hence, the different magnitudes in the shifts for the aforementioned
electropositive and electronegative adsorbates are merely
a consequence of the different type of interaction with the surface atoms,
irrespective of the applicability of any underlying charge transfer concept.
As a conclusion, we point out that SCLSs certainly are a sensitive probe of
the local electronic structure around an atom, yet they intricately depend on
the details of the interaction present in the system, which has to be properly
analyzed for each specific case to understand the observed shifts.
Therefore it does not make much sense to compare magnitudes of SCLSs arising in
chemically different systems. On the other hand, within one type of chemistry,
as e.g. in our case with the same adsorbate on the same substrate only at
different coverages, the SCLSs may indeed be used to further describe the
bonding situation -- even in the more conceptual language of charge transfer.

\subsection{O on Ru(0001) and Rh(111)}

In this view, the equal spacing of $\sim 400$ meV between SCLSs of 
increasingly higher O-coordinated Ru atoms ($S_1$, $S_1(\mbox{1O})$, $S_1(\mbox{2O})$,
and $S_1(\mbox{3O})$) suggests that the type of bonding remains the same
throughout the whole coverage range studied, or in other words, that the
(unspecified) amount of charge transferred to each O atom remains
approximately constant. This interpretation is corroborated by an almost
unchanged O${}_{1s}$ core level position to within $\pm 20$ meV. In
particular there is no indication of a qualitatively different chemisorption
behaviour between the low coverage ( $p(2\times 2)$ and $p(2\times 1)$ )
and the high coverage ( $(2\times 2)$-3O and $(1\times 1)$-O ) phases, that
could explain the long-time believed, but only apparent saturation coverage
of $\Theta = 0.5$ ML in UHV \cite{pfnuer89,madey75}. As was already concluded
in previous studies, this saturation arises therefore solely by kinetic
hindrance of the O${}_2$ dissociation process \cite{stampfl96a,stampfl96b}.
Note, that a similar picture was derived in a recent experimental study
on the O/W(110) system, which also exhibited O-coordination dependent
SCLSs up to $\approx$ 1 eV for the threefold coordinated W atoms
\cite{riffe98}.

Apart from this large scale trend, the SCLSs reflect also more subtle
details of the bonding situation. This can be seen in the differences in
the shifts for equally coordinated atoms present at two coverages;
e.g. the shifts for the $S_1(\rm{1O})$ type atoms in either the
$p(2 \times 2)$ or the $p(2\times 1)$ phase differ by 57 meV (cf. Table II).
These small variations can be due to a small redistribution of the charge
at the two coverages, which one may interpret as a slightly different
ionicity of the bond caused by the increased repulsion in a denser
adsorbate mesh \cite{ganduglia00}. Alternatively, they could be caused
by the small differences in the atomic geometries of the two phases. In
order to develop a feeling for the separate magnitudes of these two,
interrelated effects, we calculated also the SCLSs at all coverages
for an artificial bulk truncated Ru surface with the increasing number of
O atoms always sitting in hcp sites at a fixed height corresponding to
the one we deduced for the $p(2\times 2)$ relaxed geometry. The related
shifts are stated in Table II, indicating that the geometrical changes
induced by the adsorbate do amount to small shifts up to about 90 meV.
Still, the differences between equally coordinated Ru atoms (now in
completely identical nearest-neighbour surroundings for both phases)
remain of the same order as before, reflecting now solely the slight
charge rearrangement caused by the different adsorbate mesh at the two
coverages. In this respect we further note, that this sensitivity of
the SCLSs to geometrical differences can also be used to ascertain e.g.
the adsorption site. The calculated $\Delta_{\rm{SCLS}}^{\rm{total}}$
for O in fcc sites on the surface differ by $\approx 100-200$ meV from the
ones shown in Table II and are always far outside the experimental error
bar. The $S_1(\rm{3O})$ shift of a $(1 \times 1)$-O fcc phase would e.g.
be at +718 meV. If there was a significant amount of O sitting in fcc
sites at this coverage, it would certainly show up as a shoulder in the
experimental spectrum. That this is not the case, cf. Fig. 2, proves
that the experimental $(1 \times 1)$-O phase is nearly perfect hcp,
despite the small binding energy difference between both hollow
sites \cite{stampfl96a,stampfl96b}.

\begin{figure}
       \epsfxsize=0.47\textwidth \centerline{\epsfbox{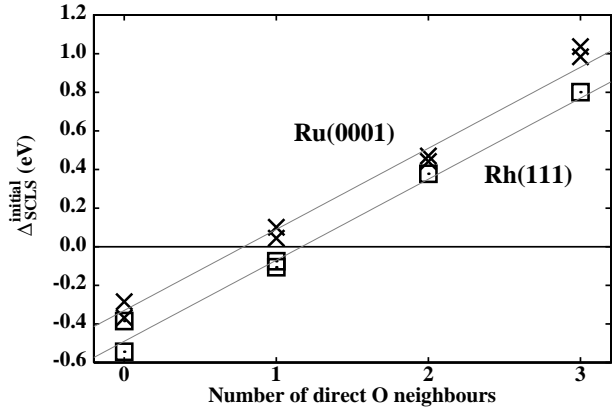}}
        \caption{Comparison of the initial state shifts,
        $\Delta_{\rm SCLS}^{\rm{initial}}$, for O/Ru(0001) (crosses)
        and O/Rh(111) (boxes) as a function
        of the number of directly coordinated O atoms. The
        lines are drawn to guide the eye. The O/Rh(111) data
        is taken from \cite{ganduglia00}.
        \label{fig:10}}        
\end{figure}

Finally, it is interesting to compare the O/Ru(0001) SCLSs to the ones found
for O/Rh(111) (same adsorbate, similar transition metal substrates)
\cite{ganduglia00}. Fig. 10 displays the calculated initial state shifts
sorted according to the number of directly coordinated O atoms. Apart from
the different SCLSs of the clean surfaces caused by the different $4d$-band
filling \cite{spanjaard85}, it is immediately obvious that both materials
display almost the same relative O-induced shifts in the whole coverage
range considered. The conclusion from these data is hence in line with the
one of preceding DFT studies concerning the adsorption energetics
\cite{stampfl96b,ganduglia99}, which apart from the different adsorption
site (hcp and fcc on Ru(0001) and Rh(111) respectively) found no qualitative
difference in the on-surface O chemisorption behaviour. In particular, in
this coverage range there is no hint towards a different catalytic behaviour
of both materials at higher O partial pressures \cite{peden86,peden88},
which hence presumably arises from different oxidation characteristics only
after O has started to penetrate into the sample. As a preliminary result
from on-going studies concerning this regime, we would like to mention that
in contrast to its near constancy in on-surface O phases, we find the
O${}_{1s}$ core level to be particularly sensitive to variations in the
sub-surface O coverage and geometrical position. This suggests that future
experimental studies dedicated to sub-surface O and surface oxide formation
should focus on this core level, rather than on the metal $3d$, which we find
to somehow saturate at its $(1 \times 1)$-O value.

\section{Summary}

SCLS experiments have been performed on the clean Ru(0001) surface and 
on the four oxygen ordered adlayer structures which form in UHV, namely the
$p(2\times 2)$, $p(2\times 1)$, $(2\times 2)$-3O and $(1\times 1)$-O.
For the clean surface the high energy resolution photoelectron diffraction
approach was used in order to make the assignment of the shifts measured to
the corresponding substrate atoms. For the oxygen related SCLSs we
find a clear dependence of the SCLS on the number of nearest neighbour O atoms,
with the higher O coordinated Ru atoms exhibiting shifts up to 1 eV
to higher binding energies.
We obtain very good agreement between the experimentally determined SCLSs
and first principles calculations, which confirms that within the GGA the
latter are able to describe this quantity with high accuracy ($\pm 30$ meV).
Using the theoretical approach, it was possible to separate the total
SCLSs into initial and final state contributions. We found the latter
to be mainly due to an enhanced intra-atomic $4d$-electron screening at
the surface, which arises from the increased $4d$-DOS at and above the
Fermi level compared to the bulk situation. The initial state shifts are
connected to a varying width of the Ru valence $4d$ band either due to
the reduced coordination of the atoms at the surface or to 
the interaction with the O $2p$ level which causes the formation of
bonding and antibonding states widening the band.
As the width of the band is connected to the formation
of bonds, which scale with the number of directly bound O atoms,
similar SCLSs result for equally O coordinated Ru atoms.
The almost linear increase of $\Delta_{\rm{SCLS}}^{\rm{initial}}$ for
increasingly higher O coordinated Ru atoms suggests that the type of
bonding remains roughly the same over the considered coverage sequence
up to the full monolayer, which may be interpreted as an almost constant
amount of charge transferred to each electronegative O atom. This
finding is similar to the result for O on Rh(111) \cite{ganduglia00},
i.e. both surfaces show a qualitatively similar on-surface chemisorption
behaviour. On the other hand the screening properties of both surfaces
are different in that the Ru(0001) surface is always able to screen the
created core hole better than the bulk while the Rh(111) surface 
screens better only for the low coverage O phases.

 These results show that a combined
experimental and theoretical determination of SCLSs provides valuable
insight into the O-metal interaction in different chemical environments.
Hence, SCLSs offer a promising tool to study not only the on-surface
O chemisorption behaviour of surfaces, but also the transition to
sub-surface O and surface oxide formation.

\begin{appendix}
\section{}

The fit of the experimental data was performed in two different ways, 
named strategy (i) and strategy (ii). The line shape parameters of
the various components are the Lorentzian and Gaussian width, L and
G in eV respectively, as well as the asymmetry parameter $\alpha$.

\subsection{strategy (i)}

The fitting procedure strategy (i) is completely independent of the
theoretical results and assumes the $S_2$(1O) component to be
indistinguishable from the bulk in all the fits. This assumption
rests on the spectrum for the $(1\times 1)$-O phase, where the
bulk and $S_1$(3O) are far from each other and the clear-cut
two peak spectrum with small overlap in between does not justify
a third component hidden under either peak at first glance, cf. Fig. 2.

The approach used to fit the data was the following:

1) First the $(1\times 1)$-O structure was fitted, for which only two components 
were assumed to be present which must be bulk and the $S_1$(3O). In this way
we found the line shape parameters of the bulk (L = 0.175, $\alpha$ = 0.085,
G = 0.11) and the $S_1$(3O) (L = 0.31, $\alpha$ = 0.150, G = 0.11) peaks.

2) Then we fitted the clean surface. In this case three components are present:
$S_1$, $S_2$ and bulk. We kept the asymmetry parameter and the Lorentzian width
for all components at the values found previously for the bulk in the $(1\times 1)$-O,
and we let free the Gaussian width of the $S_2$ and $S_1$. The Gaussian width of
$S_2$ turns out to be 0.11 eV, the same as for the bulk, while that of $S_1$ is 
0.13 eV. The assignment to first and second layer atoms, shown in Fig. 2, has
been corroborated by independent SCLS-photoelectron diffraction experiments as 
described in section IV B.

3) Next we fitted the spectra at 352 eV of the $p(2\times 2)$ in order to
determine the parameters of the $S_1$(1O) peak (L = 0.30, $\alpha$ = 0.085, G = 0.11).
These parameters are not as accurate because of the strong overlap of this peak
with that of the bulk and the other peaks present.

4) Then we fitted the $p(2\times 1)$ spectrum in order to determine the 
parameters of the $S_1$(2O) component (L = 0.30, $\alpha$ = 0.085, G = 0.11).
The parameters for this peak are not as accurate as for the $S_1$(3O), but are
definitely more accurate than those of the $S_1$(1O).

5) Finally we fitted the $(2\times 1)$3O using the line shape parameters
found previously for the various components.

\subsection{strategy (ii)}

1) In the second strategy the clean surface was fitted first. In the fit we kept
the Lorentzian width the same for the three components, letting free the
asymmetry and the Gaussian width. Fitted this way, the Lorentzian width is
0.18, the asymmetry turns out to be the same for all components, 0.086,
and the Gaussian width of the $S_1$, $S_2$ and bulk peak is 0.13, 0.09,
and 0.08 respectively. The quality of the fit was slightly better than that
of the fit of the clean surface using the first strategy, while the 
derived SCLSs were almost the same: $S_1 = -360$ meV and $S_2 = +127$ meV.

2)Then we tried to fit the $(1\times 1)$-O structure fixing for the bulk 
peak the same line shape parameters found for the clean surface and assuming
that only two components, bulk and $S_1$(3O), are present. In line with the
theoretical prediction, the bad quality of the fit rendered it necessary to
fix a third non-zero component, $S_2$(1O), at slightly lower binding energy
than the bulk peak. We fixed for this new peak the same line shape parameters
as for the bulk. By fitting the $(1\times 1)$-O structure with these three
peaks instead of two, the parameters of the $S_1$(3O) do not change with respect 
to the first fitting strategy. The bulk and the $S_2$(1O) components show
similar intensities. The SCLS for $S_1$(3O) and $S_2$(1O) turn out at 920 meV
and -60 meV respectively, both now in excellent agreement with the theoretical
values.

3) Similarly, we tried to add a non-zero $S_2$(1O) peak close to the bulk
region for all other structures, but the results were meaningless since
too many peaks are present in a small energy range.

\end{appendix}


\begin{references}

\bibitem{bormet94}
J. Bormet, J. Neugebauer, and M. Scheffler, Phys. Rev. B {\bf 49}, 17242 (1994).

\bibitem{scheffler00}
M. Scheffler and C. Stampfl, \emph{Theory of Adsorption on Metal
Substrates}, In: Handbook of Surface Science, Vol. 2: Electronic Structure,
(Eds.) K. Horn, M. Scheffler. Elsevier Science, Amsterdam (2000), p. 286.

\bibitem{mulliken55}
R.S. Mulliken, J. Chem. Phys. {\bf 23}, 1833 (1955).

\bibitem{hoffmann88}
R. Hoffmann, \emph{Solids and Surfaces: A Chemist's View of Bonding in
Extended Structures}, VCH, New York (1988).

\bibitem{bader90}
R.F.W. Bader, \emph{Atoms in Molecules. A Quantum Theory}, Int. Series
of Monographs on Chemistry, Vol. 22, Oxford University Press, Oxford (1990).

\bibitem{becke90}
A.D. Becke and K.E. Edgecombe, J. Chem. Phys. {\bf 92}, 5397 (1990).

\bibitem{martensson94}
N. M{\aa}rtensson and A. Nilsson, \emph{High Resolution Core-Level Photoelectron 
Spectroscopy of Surfaces and Adsorbates}, Springer Series in Surface Science, 
Vol. 35, (1994) 65.

\bibitem{baraldi00a}
A. Baraldi, S. Lizzit, and G. Paolucci, Surf. Sci. {\bf 457}, L354 (2000).

\bibitem{spanjaard85}
D. Spanjaard, C. Guillot, M.C. Desjonqueres, G. Treglia, and J. Lecante, 
Surf. Sci. Rep. {\bf 5}, 1 (1985); W. F. Egelhoff, Surf. Sci. Rep. {\bf 6},
253 (1987).

\bibitem{andersen94}
J.N. Andersen, D. Hennig, E. Lundgren, M. Methfessel, R. Nyholm, and
M. Scheffler, Phys. Rev. B {\bf 50}, 17525 (1994).

\bibitem{ganduglia00}
M.V. Ganduglia-Pirovano, M. Scheffler, A. Baraldi, S. Lizzit, G. Comelli,
G. Paolucci, and R. Rosei, Phys. Rev. B, {\em submitted}.

\bibitem{lindroos89}
M. Lindroos, H. Pfn\"ur, G. Held, and D. Menzel, Surf. Sci. {\bf 222},
451 (1989).

\bibitem{pfnuer89}
H. Pfn\"ur, G. Held, M. Lindroos, and D. Menzel, Surf. Sci. {\bf 220},
43 (1989).

\bibitem{gsell98}
M. Gsell, M. Stichler, P. Jakob, and D. Menzel, Israel J. Chem. {\bf 38},
339 (1998).

\bibitem{stampfl96a}
C. Stampfl, S. Schwegmann, H. Over, M. Scheffler, and G. Ertl,
Phys. Rev. Lett. {\bf 77}, 3371 (1996).

\bibitem{stampfl96b}
C. Stampfl and M. Scheffler, Phys. Rev. B {\bf 54}, 2868 (1996).

\bibitem{lizzit98}
S. Lizzit, K. Pohl, A. Baraldi, G. Comelli, V. Fritzsche, E. W. 
Plummer, R. Stumpf, and Ph. Hofmann, Phys. Rev. Lett. {\bf 81}, 3281 (1998).

\bibitem{baraldi00}
A. Baraldi, S. Lizzit, G. Comelli, A. Goldoni, Ph. Hoffmann, and
G. Paolucci, Phys. Rev. B {\bf 61}, 4534 (2000).

\bibitem{baraldi95}
A. Baraldi, M. Barnaba, B. Brena, D. Cocco, G. Comelli, S. Lizzit, G. 
Paolucci, and R. Rosei, J. Electr. Spectroscopy {\bf 76}, 145 (1995).

\bibitem{baraldi94}
A. Baraldi and V. R. Dhanak, J. Electr. Spectroscopy {\bf 67}, 211 (1994).

\bibitem{gori99}
L. Gori, R. Tommasini, G. Cautero, D. Giuressi, M. Barnaba, A. Accardo,
S. Carrato, and G. Paolucci, Nucl. Instr. and Meth. A {\bf 431}, 338 (1999).

\bibitem{stichler}
M. Stichler, Ph.D. thesis, Techn. Universit\"at M\"unchen (1998).

\bibitem{perdew96}
J.P. Perdew, S. Burke and M. Ernzerhof, Phys. Rev. Lett. {\bf 77},
3865 (1996).

\bibitem{blaha99}
P. Blaha, K. Schwarz and J. Luitz, {\bf WIEN97}, \emph{A Full Potential
Linearized Augmented Plane Wave Package for Calculating Crystal
Properties}, Karlheinz Schwarz, Techn. Universit\"at Wien, Austria,
(1999). ISBN 3-9501031-0-4.

\bibitem{kohler96}
B. Kohler, S. Wilke, M. Scheffler, R. Kouba, and C. Ambrosch-Draxl,
Comput. Phys. Commun. {\bf 94}, 31 (1996).

\bibitem{petersen00}
M. Petersen, F. Wagner, L. Hufnagel, M. Scheffler, P. Blaha, and
K. Schwarz, Comp. Phys. Commun. {\bf 126}, 294 (2000).

\bibitem{johansson80}
B. Johansson and N. M{\aa}rtensson, Phys. Rev. B {\bf 21}, 4427 (1980).

\bibitem{janak78}
J.F. Janak, Phys. Rev. B {\bf 18}, 7165 (1978); J.P. Perdew and
M. Levy, Phys. Rev. B {\bf 56}, 16021 (1997).

\bibitem{doniach70}
S. Doniach and M. $\breve{\rm{S}}$\`unji\'c, J. Phys. C {\bf 3}, 185 (1970).

\bibitem{woodruff94}
D.P. Woodruff and A.M. Bradshaw, Rep. Prog. Phys. {\bf 57}, 1029 (1994), and 
references therein.

\bibitem{vanhove}
Y. Chen and M.A. Van Hove, http://electron.lbl.gov/mscdpack/.

\bibitem{rehralbers}
J.J. Rehr and R.C. Albers, Phys. Rev. B {\bf 41}, 8139 (1990).

\bibitem{over92}
H. Over, H. Bludau, M. Skottke-Klein, G. Ertl, W. Moritz, and C.T. 
Campbell, Phys. Rev. B {\bf 45}, 8638 (1992).

\bibitem{perdew92}
J.P. Perdew and Y. Wang, Phys. Rev. B {\bf 45}, 13244 (1992).

\bibitem{stampfl01}
C. Stampfl, M.V. Ganduglia-Pirovano, K. Reuter, and M. Scheffler,
Surf. Sci. {\bf 500}, {\em submitted}.

\bibitem{methfessel93}
M. Methfessel, D. Hennig, and M. Scheffler, Surf. Sci. {\bf 287/288},
785 (1993).

\bibitem{methfessel95}
M. Methfessel, D. Hennig, and M. Scheffler, Surf. Rev. Lett. {\bf 2},
197 (1995).

\bibitem{friedel69}
J. Friedel, In: \emph{The Physics of Metals}, (Ed.) J.M. Ziman,
Cambridge University Press, Cambridge (1969), p. 340.

\bibitem{remark1}
We calculate the $4d$-DOS, $N_{4d}(\epsilon)$, only inside the muffin
tin spheres of the FP-LAPW scheme. As can be seen in Fig. 9, the $4d$
orbital extends beyond the currently chosen sphere of 2.3 Bohr radius
(equivalent to 91\% of half the nearest neighbour distance in bulk Ru).
Yet, these tails of the $4d$ orbital hardly affect the shape of 
$N_{4d}(\epsilon)$, which is what we are primarily interested in.
We checked this by reducing the muffin tin radius by 10\%. While the
calculated $4d$-DOSs simply contained fewer states, their overall
shape was almost unchanged. This is reflected in the computed moments
of the DOS, e.g. for the bulk Ru $4d$-band, where the number of
states, $\mu_0$, changes from 7.85 (2.3 bohr) to 6.97 (2.0 bohr), but
the center of gravity, $\epsilon_c$, and the width, $W$, change only
by 60 meV and 0.2\% respectively. Note particularly that our intention
in presenting the $4d$-DOSs is not to make a quantitative theory, but
'only' to explain our DFT results conceptually.

\bibitem{alden93}
M. Ald{\'e}n , H.L. Skriver, and B. Johansson, Phys. Rev. Lett. {\bf 71},
2449 (1993); M. Ald{\'e}n, I.A. Abrikosov, B. Johannson, N.M. Rosengaard,
and H.L. Skriver, Phys. Rev. B {\bf 50}, 5131 (1994).

\bibitem{pettifor95}
D. Pettifor, \emph{Bonding and Structure of Molecules and Solids},
Clarendon Press, Oxford (1995).

\bibitem{remark2}
As apparent from Fig. 7, the upper integration limit in eq. (4) is
not well defined. We integrated all states up to
$\epsilon_{up} = \epsilon_F + 2.5$ eV, which is at all considered coverages
above the strong decrease in the $d$-DOS at the top of the band (cf. Fig.
7). While the absolute moments, $\mu_p$, certainly depend on the choice
of this upper integration limit, the relative changes in the moments,
$\Delta \mu_p = \mu_p(\Theta_1) - \mu_p(\Theta_2)$, for different
coverages are not very sensitive to it. We tested this by varying the
upper integration limit to either $\epsilon_{up} = \epsilon_F + 2.0$ eV or
$\epsilon_{up} = \epsilon_F + 3.0$ eV. The changes in $\Delta C_{4d}$
were within $\pm 60$ meV and the $\Delta W$s differed by less than 2\%.
These variations are small compared to the overall trend which we intend
to explain, so that the conclusions drawn with the present limit are
independent of the particular choice made.

\bibitem{remark3}
The values for the rectangular $d$-band model of bulk Ru are determined
as: bottom of band, $\epsilon_{dn} = -5.81$ eV, top of band, 
$\epsilon_{up} = +2.49$ eV, center of gravity, $\epsilon_c = -1.66$ eV,
and total band width, $W = 8.30$ eV. The band filling is 70\% and the
energy zero is the Fermi level.

\bibitem{ganduglia96}
M.V. Ganduglia-Pirovano, V. Natoli, M.H. Cohen, J. Kudrnovsk{\'y}, and
I. Turek, Phys. Rev. B {\bf 54}, 8892 (1996).

\bibitem{benesh92}
G.A. Benesh and D.A. King, Chem. Phys. Lett. {\bf 191}, 315 (1992).

\bibitem{bagus93}
P.S. Bagus, C.R. Brundle, G. Pacchioni, and F. Parmigiani,
Surf. Sci. Rep. {\bf 19}, 266 (1993).

\bibitem{bagus99}
P.S. Bagus, F. Illas, G. Pacchioni, and F. Parmigiani, J. Electr. Spectroscopy
{\bf 100}, 215 (1999).

\bibitem{riffe90}
D.M. Riffe, G.K. Wertheim, and P.H. Citrin, Phys. Rev. Lett. {\bf 64},
571 (1990).

\bibitem{madey75}
T.E. Madey, H.A. Engelhardt, and D. Menzel, Surf. Sci. {\bf 48},
304 (1975).

\bibitem{riffe98}
D.M. Riffe and G.K. Wertheim, Surf. Sci. {\bf 399}, 248 (1999).

\bibitem{ganduglia99}
M.V. Ganduglia-Pirovano and M. Scheffler, Phys. Rev. B {\bf 59},
15533 (1999).

\bibitem{peden86}
C.H.F. Peden and D.W. Goodman, J. Phys. Chem. {\bf 90}, 1360 (1986).

\bibitem{peden88}
C.H.F. Peden, D.W. Goodman, D.S. Blair, P.J. Berlowitz, G.B. Fischer,
and S.H. Oh, J. Phys. Chem. {\bf 92}, 1563 (1988). 

\end{references}
\end{document}